\documentclass[aps,twocolumn,showpacs,pre,10pt]{revtex4-2}
\usepackage{epsfig,amsmath,color,colordvi,verbatim}
\usepackage[normalem]{ulem}
\usepackage{braket}
\usepackage{float}
\usepackage{hyperref}

\def\be{\begin{equation}}
\def\ee{\end{equation}}
\def\bee{\begin{eqnarray}}
\def\eee{\end{eqnarray}}

\begin{document}
%\title{Controlling the quantum yield of photoisomerization}
%\title{Vibrational coherence of an electronic wave packet at a conical
%intersection}
\title{Quantum Dynamics of Enantiomers in Chiral Optical Cavities} 

\author{Yang-Cheng Ye$^{1,*}$}
\author{Panpan Zhang$^{1,*}$}
\author{Ajay Jha$^{2,3}$}\email{Ajay.Jha@rfi.ac.uk}
\author{Fulu Zheng$^{1}$}\email{zhengfulu@nbu.edu.cn}
\author{Hong-Guang Duan$^{1}$}\email{duanhongguang@nbu.edu.cn} 
\affiliation{
$^1$Department of Physics, School of Physical Science and Technology, Ningbo University, Ningbo, 315211, P.R. China \\
$^2$Rosalind Franklin Institute, Harwell, Oxfordshire OX11 0QX, United Kingdom\\
$^3$Department of Pharmacology, University of Oxford, Oxford, OX1 3QT United Kingdom 
}
\date{\today}

\begin{abstract} 

Chirality, the absence of mirror symmetry, is a fundamental molecular property with far-reaching consequences from chemistry to biology. Yet enantiosensitive optical responses are very weak. Here, we introduce a theoretical framework in which a chiral optical cavity under strong coupling directly lifts the degeneracy of opposite enantiomers at the electronic-dipole level. The cavity’s parity-breaking field inside the cavity induces distinct site-energy shifts for left- versus right-handed molecules, producing robust enantioselective polariton states that overcome the weakness of traditional chiroptical effects. Using cavity quantum electrodynamics simulations, we show that strong light-matter coupling reshapes the polaritonic energy landscape and leads to enantiomer-specific coherence lifetimes and relaxation pathways. To reveal these dynamics, we propose ultrafast two-dimensional electronic spectroscopy (2DES) as a probe, capable of resolving polaritonic splittings on femtosecond timescales. Simulated 2DES spectra exhibit unambiguous enantioselective signatures of the cavity-induced asymmetry. These findings establish that chiral cavities provide a powerful platform for detecting and controlling molecular handedness beyond the limits of conventional optical methods. 

\end{abstract}

\maketitle

\maketitle

% ----------- Introduction -------------------------

Chirality, the lack of mirror symmetry, is a fundamental property of matter with profound consequences across physics, chemistry, and biology \cite{ref1, ref2, ref3, ref4}. Enantiomers (non-superimposable mirror images) are energetically identical in free space, yet they respond differently to electromagnetic fields. These distinctions, while subtle, underlie diverse phenomena from molecular recognition to the handedness of biological systems \cite{ref5, ref6, ref7}. A long-standing challenge is to resolve enantiomer-specific dynamics with sufficient strength to allow for unambiguous discrimination.

Conventional strategies employ chiral light, such as circularly polarized beams, to distinguish enantiomers through signals like circular dichroism \cite{ref8, ref9}. However, such observables rely on higher-order multipolar couplings (magnetic dipoles or electric quadrupoles), which are typically suppressed by factors of 10$^{-2}$ to 10$^{-3}$ relative to electric-dipole interactions. As a result, chiral light-matter couplings are intrinsically weak, limiting both sensitivity and control. This motivates the search for mechanisms that can enhance enantioselective interactions at the fundamental level.

Optical cavities provide a natural solution \cite{ref10}. By confining photons in high-Q resonators with sub-wavelength mode volumes, cavity quantum electrodynamics (QED) dramatically amplifies light-matter coupling \cite{ref11, ref12, ref13}. In the strong-coupling regime, molecular excitations hybridize with cavity photons to form polaritons, which is mixed light-matter quasiparticles with properties distinct from either constituent \cite{ref14, ref15, ref16, ref17, ref18, ref18a}. Importantly, even the electromagnetic vacuum inside the cavity reshapes molecular spectra, shifting transition energies and altering relaxation channels \cite{ref19, ref20, ref21}. When cavity modes are engineered to break mirror symmetry, they provide an entirely new route to enantiomer discrimination \cite{ref22}. Unlike free-space chiral light, where handedness enters only through weak multipolar corrections, chiral cavities generate enantioselective shifts already at the dipolar level \cite{ref23, ref24}. Left- and right-handed molecules thus acquire different site energies and experience distinct polaritonic potential landscapes \cite{ref25}. This mechanism circumvents the weakness of traditional chiral observables and offers a controllable platform for enantioselective quantum dynamics.
The underlying physics can be described within standard cavity QED \cite{ref26}. Chiral cavities introduce spatial asymmetries in the quantized field that break parity and lift the degeneracy of enantiomers. The resulting dynamics evolve on distinct polaritonic energy surfaces, enabling control over coherence lifetimes, population transfer, and nonadiabatic processes with enantioselective resolution. A theoretical description of these effects requires going beyond perturbation theory. In the strong-coupling regime, coherent exchange between light and matter competes with dissipative channels, necessitating frameworks such as polaritonic master equations or path-integral approaches \cite{ref27}. These methods capture the interplay of coherence, cavity-induced asymmetry, and environmental decoherence that governs enantiomer dynamics inside resonators.

The challenge then becomes probing such dynamics experimentally. Enantiomer-specific linear spectroscopic signals vanish in racemic mixtures and are often buried beneath achiral backgrounds. Nonlinear multidimensional methods, by contrast, can isolate weak signals and reveal coherent dynamics directly \cite{ref28, ref29}. Two-dimensional electronic spectroscopy (2DES) is especially powerful \cite{ref30, ref31}. By correlating excitation and detection frequencies on femtosecond timescales, 2DES distinguishes coherent oscillations from population relaxation, resolves polaritonic splitting, and uncovers couplings inaccessible to linear probes \cite{ref31a, ref32, ref33, ref34}. In chiral cavities, these capabilities translate into clear enantioselective signatures, visible through cross-peaks and phase-sensitive detection schemes. Equally important, 2DES can track nonadiabatic processes in real time. Strong coupling has been shown to shift conical intersections and suppress internal conversion, modifying ultrafast relaxation pathways \cite{ref19, ref35}. These modifications manifest as distinct changes in coherence lifetimes and spectral branching ratios. In the presence of chiral cavity fields, opposite enantiomers follow different polaritonic trajectories, producing unique multidimensional fingerprints that enable discrimination even in mixed ensembles.

In this paper,  we present a theoretical framework in which a chiral optical cavity modifies enantiomer dynamics via parity breaking at the electronic-dipole level. This novel light-matter interaction yields the complex value, which drives the enantiomers-distincted optical transitions. It produces distinct polaritonic energy shifts and relaxation dynamics for left- and right-handed molecules. 2DES reveals these effects through shifted diagonal peaks, differential cross-peaks, and time-resolved coherence oscillations. The results demonstrate that cavity chirality imprints measurable asymmetries on ultrafast dynamics, establishing a viable route for detecting and influencing molecular handedness through engineered light-matter interactions.

% ---------------------- Section III: Theoretical Model

\begin{figure}[t!]
\begin{center}
\includegraphics[width=8.5cm]{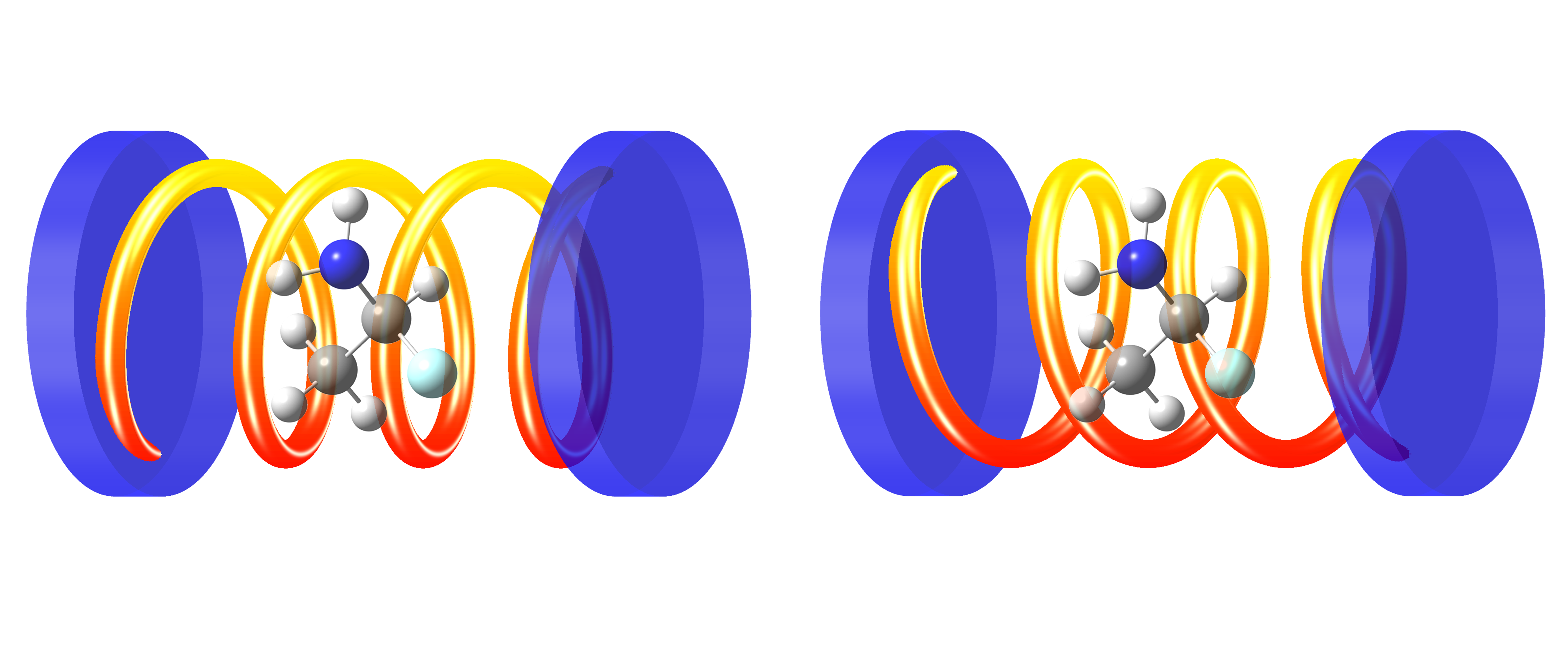}
\caption{\label{fig:Fig1}Chiral cavity (left and right circularly polarized light) with 1-Fluoroethanamine molecule inside. } 
\end{center}
\end{figure}

{\em Throretical Model:} The total Hamiltonian can be written as $H=H_s+H_b+H_{sb}$, where $H_s$ is the Hamiltonian of system and $H_b=\sum_n\hbar\overline{\omega}_n(\overline{p}^2_n+\overline{q}^2_n)$ is the bath part, which contains infinite harmonic oscillators. Here, $\overline{p}_n$ and $\overline{q}_n$ are the dimensionless momentum and coordinate operators of n-th harmonic oscillator with corresponding frequency $\overline{\omega}_n$. $H_{sb}=\sum_{ni}|i\rangle \langle i|\overline{q}_n$ describes the interaction of system and heat bath, where $\ket{i}$ are the quantum states of system. The system Hamiltonian of photons, matter and their interaction in atomic units are represented as
\begin{align}
H_s=&\frac{1}{2}\sum_i [\textbf{p}_i-\textbf{A}(\textbf{r}_i,t)]^2+\sum_I \frac{1}{2M_I}[\textbf{P}_I+Z_I\textbf{A}(\textbf{R}_I,t)]^2\nonumber \\
	&+\sum_{i>j}\frac{1}{|\textbf{r}_i-\textbf{r}_j|}+\sum_{I>J}\frac{Z_IZ_J}{|\textbf{R}_I-\textbf{R}_J|}-\sum_{i,I}\frac{Z_I}{\textbf{R}_I-\textbf{r}_i}\nonumber \\
	&+\frac{\epsilon_0}{2}\int (\textbf{E}^2(\textbf{r},t)+c^2\textbf{B}^2(\textbf{r},t))dr^3, \label{hs1}
\end{align}
where $i$ and $j$ label electrons while $I$ and $J$ label nuclei with charges $Z_I$ and $Z_J$. In the following, vectors are placed in bold type. $\textbf{p}_i$ and $\textbf{P}_I$ are momentum of electron $i$ and nuclei $I$, while $\textbf{r}_i$ and $\textbf{R}_I$ are coordinate of electron $i$ and nuclei $I$. The vector potential, electronic fields and magnetic fields are denoted by $\textbf{A}(\textbf{r},t)$, $\textbf{E}(\textbf{r},t)$ and $\textbf{B}(\textbf{r},t)$, respectively. $c$ is the speed of light. In strong coupling regime,  the photons are treated by quantum electrodynamics (QED) and the vector potential in Coulomb gauge can be written as  
\begin{equation}
\textbf{A}(\textbf{r},t)=\sum_k \sqrt{\frac{\hbar}{\epsilon_0 V 2\omega_k}}( \epsilon_k b_k e^{i(\textbf{k}\textbf{r}-\omega_kt)} + \epsilon^{*}_k b^{\dagger}_k e^{-i(\textbf{k}\textbf{r}-\omega_kt)}), 
\end{equation}
where $k$ contains all possible wave vectors and $\bm{\epsilon}$ denotes the field polarization. $V$ is the quantization volume inside the chiral cavity, while $\omega_k$ is the corresponding allowed frequency of light. $b^{\dagger}$ and $b$ are the creation and annihilation operators of chiral cavity. In QED, the interaction of electromagnetic field and Dirac field has no analytic result. We simply assume they have time dependence as plane waves. If the wave vector $\textbf{k}$ is assumed along the z axis and only have two possible wave vectors $(0,0,k)$ and $(0,0,-k)$, the vector potential's polarization of left and right chiral cavity is 
\begin{equation}
\bm{\epsilon}_{k}^L=\frac{1}{\sqrt{2}}\begin{pmatrix}
		1\\
		i\\
		0
\end{pmatrix},
\bm{\epsilon}_{k}^R=\frac{1}{\sqrt{2}}\begin{pmatrix}
		1\\
		-i\\
		0
\end{pmatrix}.
\end{equation}
Under the Born-Oppenheimer approximation, the Hamiltonian of nuclei is neglected. Then we use the Gaussian software (Gaussian 16 \cite{Gaussian}) to optimize the molecular structure and calculate the electronic states ($\ket{m}$) at the DFT-b3lyp using the cc-pvdz basis set. Multiwfn \cite{Multiwfn} is used to output the Gaussian type orbitals of corresponding electronic states. So the Eq.\ (\ref{hs1}) becomes 
\begin{align}
H_s=\sum_m E_m|m\rangle\langle m|-\sum_i \textbf{p}_i\cdot\textbf{A}(\textbf{r}_i,t)\nonumber \\
+\frac{N_e}{2}\textbf{A}^2(\textbf{r},t)+\omega_k (b_{k}^{\dagger}b_{k}+b_{-k}^{\dagger}b_{-k}),\label{hs2}
\end{align}
where $N_e$ accounts the number of electrons of one molecule inside the chiral cavity. Here, we introduce the coupling strength $g$ defined as $g=\sqrt{\frac{\hbar}{\epsilon_0V2\omega_k}}$. Considering the property of electromagnetic vector potential (see Sec.\ VI of the SI) and inserting the expression of $\textbf{A}$($\textbf{r}$,t), the Eq.\ (\ref{hs2}) can be rewritten as 
\begin{align}
H_s=&\sum_m E_m|m\rangle\langle m|
	+\omega_k (b_{k}^{\dagger}b_{k}+b_{-k}^{\dagger}b_{-k}) \nonumber \\
	&+\frac{N_e g^2}{2}(b_k+b_{-k}^{\dagger})(b^{\dagger}_k+b_{-k})
	\nonumber\\
	&-g\sum_i \textbf{p}_i\cdot\bm{\epsilon}_k(b_k+b_{-k}^{\dagger})e^{i(\textbf{k}\textbf{r}-\omega_kt)} \nonumber \\
	&-g\sum_i \textbf{p}_i\cdot\bm{\epsilon}^{*}_k(b^{\dagger}_k+b_{-k})e^{-i(\textbf{k}\textbf{r}-\omega_kt)}.\label{hs3}
\end{align}
As illustrated in Fig.\ \ref{fig:Fig1}, the chiral molecule 1-Fluoroethanamine with R and S enantiomers is chosen as the model molecule inside the chiral cavity. The electronic ground state $|S_0\rangle$, two excited states $|S_1\rangle$ and $|S_2\rangle$ are taken into account in the study. The energies of the three states are $E_0$ = 0 eV, $E_1$ = 6.9866 eV, $E_2$ = 9.3447 eV, respectively. The two modes of chiral cavity are set as 2-levels and frequency $\omega_k$ = 4.0 eV. The coupling strength $g$ is set as 0.05 a.u.. The schematic diagram of molecules inside chiral cavity has been illustrated in Fig.\ \ref{fig:Fig1}. 

%%%%%%%%%%%%%%%%%%%%%%%%%%%%%%%%%%%%%%%%%%%%%%%%%%
%%%%%%%%%%%%%%%%%%%%%%%%%%%%%%%%%%%%%%%%%%%%%%%%%%

% ---------------------- Section IV: Results and Discussion 
%\section*{Results and Discussion} 
 
With system Hamiltonian and its light-matter interaction terms, we employed the Redfield quantum master equation \cite{Redfield} with double-side Feynman diagram \cite{Tanimura, Mukamel book} to calculate the 2DES with delta-shape-approximated laser pulses. The detailed derivations of quantum master equation and the calculations of response function and 2DES have been described in Sec.\ IV of the supporting information (SI). 

%\subsection{Time-resolved 2DES with left- and right-handed circularly polarized cavity light} 

\begin{figure}[t!]
\begin{center}
\includegraphics[width=9.0cm]{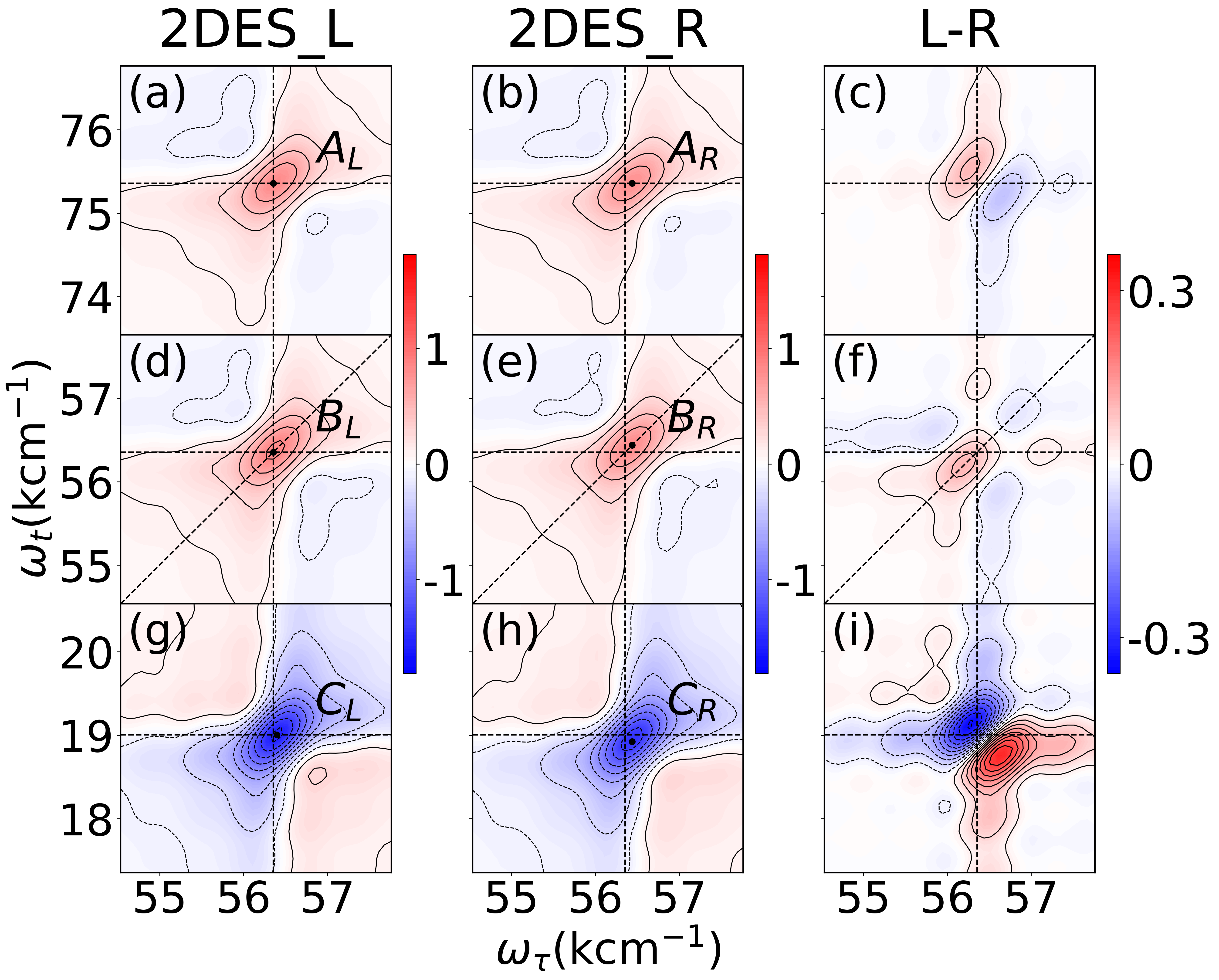}
\caption{\label{fig:Fig2}The 2DES with left- (a, d and g) and right-handed (b, e and h) chiral cavity at waiting time of T = 0 fs. The differential signal of subtracting from left- to the right-handed cavities, the resulted data is shown in (c), (f) and (i), respectively. }
\end{center}
\end{figure}

{\em 2DES at waiting time T = 0:} We firstly show the calculated 2DES at initial waiting time T =  0 fs in Fig.\ \ref{fig:Fig2}. We assumed the transition dipoles between electronically ground to two electronic excited states (S$_1$ and S$_2$). Thus, we obtained six main and cross peaks in the 2DES. The calculated total spectrum has been shown in Sec.\ X of the SI. Here, we show the resulted data with left-handed cavity light  in (a), (d) and (g) and the right-handed ones in (b), (e) and (h) in Fig.\ \ref{fig:Fig2}, respectively. Due to the complex values in light-matter interaction terms in Eq.\ \ref{hs3}, they show distinguishable energy difference between left- and right-handed parts, which has been marked by black dashed lines. Moreover, the difference of spectroscopic signals have been obtained by subtracting from left- to the right-handed parts, which has been plotted in Fig.\ \ref{fig:Fig2}(c), (f) and (i). Based on the calculations, we have shown more than 10\% of difference signal was obtained and it shows the  negative and positive peaks in Fig.\ \ref{fig:Fig2}(c). Moreover, a butterfly shape of difference result is presenting in Fig.\ \ref{fig:Fig2}(i). By this, we have demonstrated the validity of employing 2DES to distinguish the difference of site energies with left- and right-handed cavities. Moreover, the unique separation of excitation and detection windows of 2DES allow us to effectively separate main and cross peaks between two electronic excited states (S$_1$ and S$_2$). The optical assignment of red and blue peaks are described in details in the SI. The time-resolved 2DES with longer waiting times are plotted and shown in Sec.\ XI of the SI. 

The enantioselective 2DES signal exhibits a pronounced dependence on cavity detuning. As the cavity mode frequency $\omega_k$ is tuned across the electronic resonance, the magnitude and spatial distribution of the left-right differential spectra vary non-monotonically. Figures S12-S14 show that both diagonal and cross-peak asymmetries are maximised only within a narrow range of $\omega_k$, where light-matter hybridisation is strongest. Away from this resonance condition, the differential features diminish or partially invert, indicating a reorganisation of the polaritonic energy landscape. This behaviour demonstrates that cavity-induced enantioselectivity is not fixed by molecular structure alone, but can be actively enhanced or suppressed through cavity frequency control, providing a tunable handle for optimising chiral contrast in ultrafast spectroscopy.

%%%%%%%%%%%%%%%%%%%%%%%%%%%%%%%%%%%%%%%%%%%%%%%%%%
%%%%%%%%%%%%%%%%%%%%%%%%%%%%%%%%%%%%%%%%%%%%%%%%%%

\begin{figure}[t!]
\begin{center}
\includegraphics[width=9.0cm]{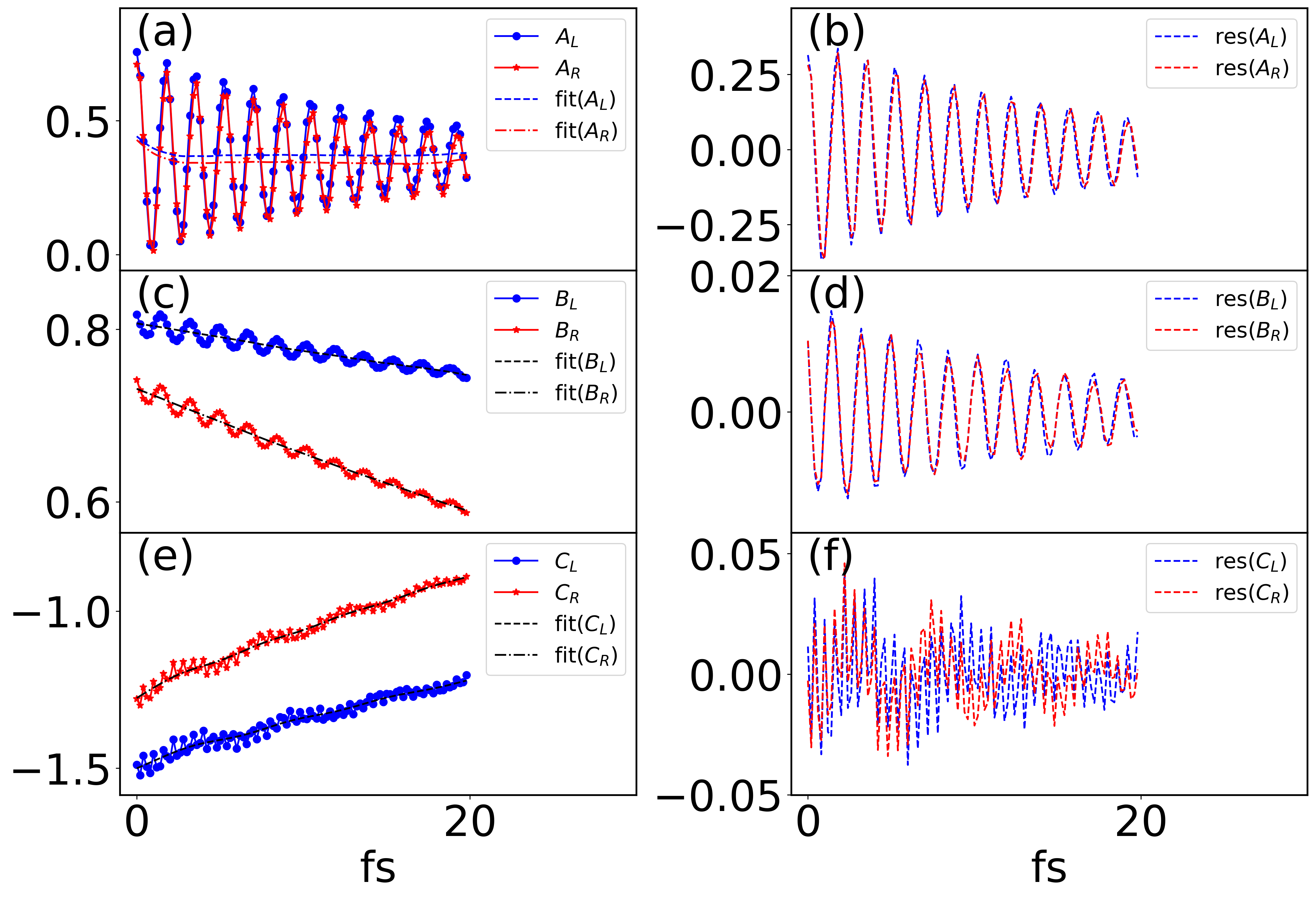}
\caption{\label{fig:Fig3}(a, c and e)Time-resolved traces and residuals of peaks A, B and C. The fitting curves are plotted as red and blue dashed lines. The time-resolved residuals are plotted as blue and red solid lines in (b), (d) and (f). }
\end{center}
\end{figure}

{\em 2DES at finite waiting time:} We then performed more calculations with different waiting time. We extracted the magnitude of main peak B and cross peak (A and C), we plotted the time-resolved amplitude in Fig.\ \ref{fig:Fig3}. The left- and right-handed cavities are plotted as blue and red dots in Fig.\ \ref{fig:Fig3}(a), (c) and (e). We observed a clear evidence of coherence, which lasts for more than 20 fs with period of 1.67 fs in Fig.\ \ref{fig:Fig3}(a). Moreover, we also observed the difference of decay timescales between red and blue lines, the fitting curves are plotted as red and blue dashed lines in Fig.\ \ref{fig:Fig3}(a). We then plotted the residuals of peak A as red and blue dashed lines in Fig.\ \ref{fig:Fig3}(b). They show excellent oscillation in phase. We also repeated the same procedure and plotted time-resolved traces of peak B in Fig.\ \ref{fig:Fig3}(c). The fitting procedure yields the different timescales of changing amplitude and the fitting curves are plotted as dashed lines. The residuals of peak B (left- and right-handed cavity) are presented in Fig.\ \ref{fig:Fig3}(d). They also present oscillation with excellent in phase. Moreover, we also plotted time-resolved traces of cross peak C as red and blue dots in Fig.\ \ref{fig:Fig3}(e). The subsequent fitting procedure showing the different decaying timescales, the fitting curves are plotted as red and blue dashed lines in Fig.\ \ref{fig:Fig3}(e). The time-resolved residuals are plotted as red and blue dashed lines in Fig.\ \ref{fig:Fig3}(f). Again, they show excellent oscillatory dynamics and in phase. 

We then performed the data analysis of 2DES with excitation of S$_2$. The resulted data of time-resolved traces are plotted in Sec.\ IX in the SI. The fitting curves are also fitted to extract the decaying lifetimes of each trace. The fitting curves are also plotted as red and blue solid lines, respectively. The time-resolved residuals are further plotted red and blue dashed lines. Again, they show clear oscillatory dynamics with oscillation in phase. The detailed results and data analysis are described in Sec.\ IX in the SI.

%%%%%%%%%%%%%%%%%%%%%%%%%%%%%%%%%%%%%%%%%%%%%%%%%%
%%%%%%%%%%%%%%%%%%%%%%%%%%%%%%%%%%%%%%%%%%%%%%%%%%

\begin{figure}[t!]
\begin{center}
\includegraphics[width=9.0cm]{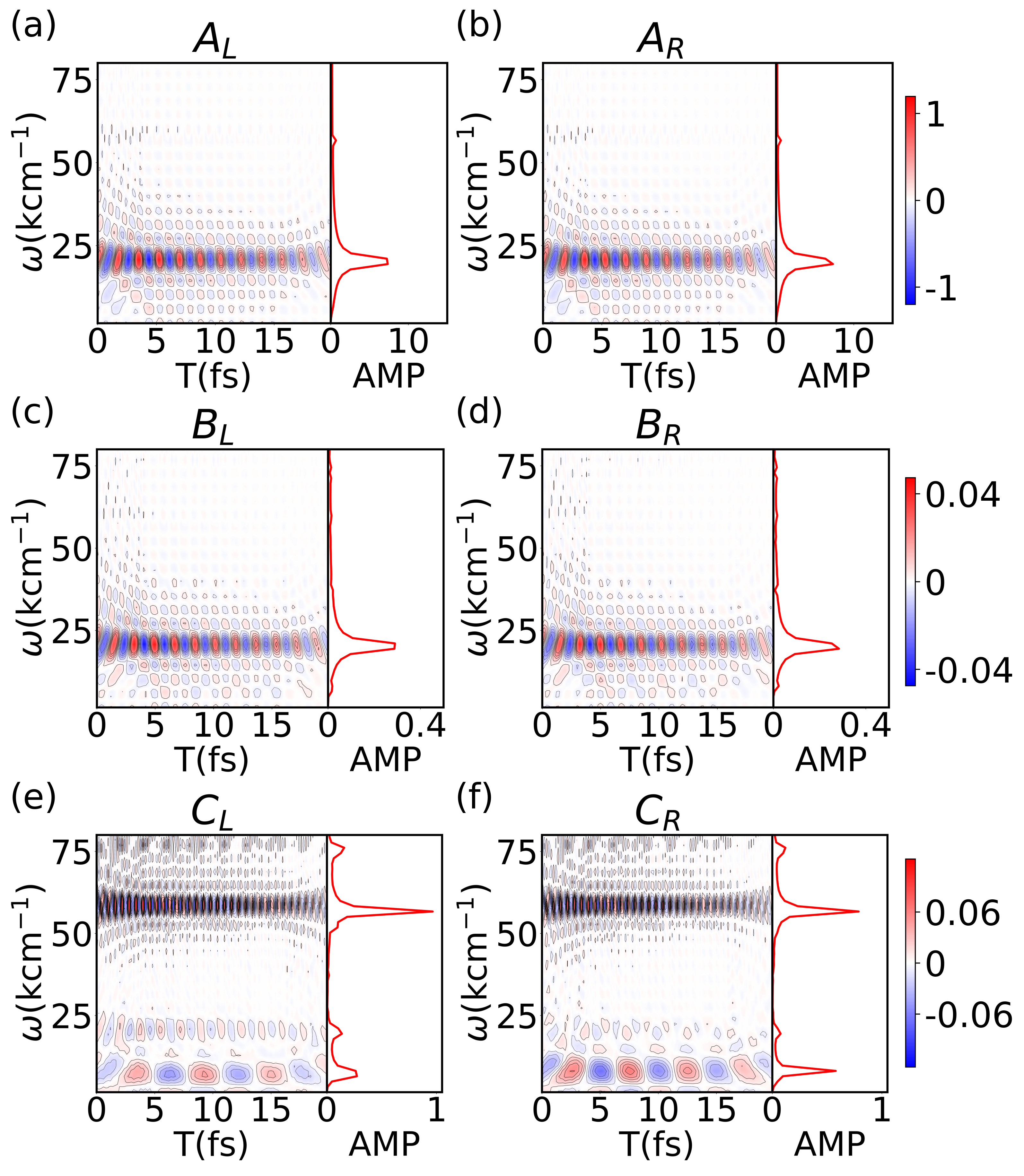}
\caption{\label{fig:Fig4}Results after wavelet analysis of peak A in (a) and (b). The wavelet resulted data of cross peak B are in (c) and (d). (c and f) show the wavelet resulted data with resolved vibrational coherences.  }
\end{center}
\end{figure}

{\em Vibrational coherence:} We then performed Fourier transform of the residuals after subtracting the kinetics by fitting procedures. We plotted the resulted data after Fourier transform in Fig.\ \ref{fig:Fig4}. Moreover, we also performed wavelet analysis of residuals and plotted the results in Fig.\ \ref{fig:Fig4} as well. We show a clear frequency at 20 kcm$^{-1}$ of oscillation in peak A (A$_{L}$ and A$_{R}$ in (a) and (b)). They show the oscillation with the same phase. Moreover, we also plotted the wavelet resulted data of peak B in (c) and (d). They show two frequencies at 20 kcm$^{-1}$ with a side peak of 10 kcm$^{-1}$. Interestingly, they also present the oscillatory dynamics with the same phase. The wavelet resulted data of peak C (C$_{L}$ and C$_{R}$) in Fig.\ \ref{fig:Fig4}(e) and (f), respectively. They show more complicated oscillatory dynamics and with more resolved vibrational modes. They include 10, 20, 60, 66 and 78 kcm$^{-1}$. In addition, the left-handed 2DES show the same phase of oscillation in Fig.\ \ref{fig:Fig4}(e) and (f). Moreover, we also presented more calculated results of oscillatory dynamics after photoexcitation of S$_2$, which is shown in Sec.\ IX in the SI. They show more clear oscillatory dynamics, the detailed description and explanation have been shown in the SI. 

Based on these calculations, we numerically demonstrated that the shift of resolved peaks in 2DES after interaction with chiral cavity is an effective approach to identify the chiral interaction between light and matter. However, we also demonstrated that time-resolved electronic or vibrational coherences can not be effectively detected and analyzed by phase difference between left- and right-handed pulses. 

%%%%%%%%%%%%%%%%%%%%%%%%%%%%%%%%%%%%%%%%%%%%%%%%%%
%%%%%%%%%%%%%%%%%%%%%%%%%%%%%%%%%%%%%%%%%%%%%%%%%%

%\section*{Discussions} 

{\em Discussion:} Under strong light-matter coupling in a chiral cavity, the 2D spectra exhibit pronounced enantioselective asymmetries. Left- and right-handed cavity conditions produce slightly different transition energies, so that corresponding peaks in the 2DES appear at shifted frequencies. This is evident for both the main (diagonal) excitonic peaks and the off-diagonal cross-peaks connecting the two excited states. Subtracting the two spectra yields differential features with characteristic positive-negative lobes (for example, a dispersive 'butterfly' pattern around certain cross-peaks) directly reflecting the small frequency offsets induced by the cavity’s chiral field. The magnitude of these spectral differences reaches over ten percent of the signal, far exceeding conventional chiroptical effects. Such pronounced contrast, clearly resolved by the 2DES technique (which isolates otherwise overlapping peaks), confirms that the cavity-induced dipolar asymmetry produces a robust enantioselective splitting at the electric-dipole level. 

Time-resolved 2DES traces further reveal long-lived coherent oscillations and enantioselective dynamics. We observe persistent quantum beats in the peak amplitudes lasting over 20 fs, with a principal period of approximately 1.7 fs (corresponding to 20 kcm$^{-1}$). This oscillation arises from the energy gap between the two excited states, manifesting as a polaritonic coherence in both enantiomeric cases. Fourier analysis also reveals additional coherence components, including a weak 10 kcm$^{-1}$ sideband to the main beat and lower-frequency oscillations (60 - 80 cm$^{-1}$) in some cross-peaks that reflect underlying vibrational modes. Importantly, these oscillations remain phase-locked between left- and right-handed scenarios (no relative phase shift is detected in the beats) implying that the chiral cavity’s effect enters through energy-level shifts rather than any initial phase bias in the coherent superposition. Beyond the coherent beats, the cavity also modulates relaxation kinetics in an enantioselective manner. By fitting the decay of each spectral feature, we find that the left- and right-handed cavity cases exhibit distinct lifetimes for the same peak. In our model system the disparity is modest (on the order of a few femtoseconds) due to the simple two-level structure, but it nonetheless demonstrates that opposite enantiomers experience different polaritonic decay pathways under chiral confinement.

These enantioselective effects are furthermore tunable by varying key cavity parameters. In the weak-coupling regime (small $g$) no discernible splitting appears, whereas increasing $g$ into the strong-coupling regime causes clear spectral differences to emerge and grow. Our simulations indicate an optimal coupling strength around $g$ = 0.05 - 0.1 (in atomic units) at which the enantiomeric energy splitting is maximal. The cavity resonance frequency provides another control knob: tuning $\omega_k$ modulates the differential signal amplitude non-monotonically, with certain $\omega_k$ values greatly enhancing the chiral response while detuned conditions diminish it. Likewise, higher cavity loss (lower $Q$) significantly suppresses the enantioselective coherence signals and reduces the disparity in relaxation times. Thus, low-loss resonant cavities in the strong-coupling regime are crucial for amplifying and sustaining the ultrafast enantiosensitive dynamics demonstrated here.

%----------------------- conclusion part ------------------
%\section*{Conclusion} 

{\em Conclusion:} In this paper, we have demonstrated theoretically that a chiral optical cavity under strong coupling can induce and probe ultrafast enantioselective molecular dynamics. In our model system, the cavity’s handed electromagnetic field lifts the degeneracy of opposite enantiomers at the electric-dipole level, giving rise to distinct polaritonic surfaces and enantio-specific energy shifts. Two-dimensional electronic spectroscopy was shown to be a powerful tool for capturing these effects, yielding clear signatures such as shifted peak positions, differential cross-peak patterns, coherence beatings, and relaxation time differences between left- and right-handed configurations. This mechanism bypasses the intrinsic weakness of conventional chiroptical signals by operating in an enhanced light-matter interaction regime, producing robust enantiomer-specific observables. Because the effect derives from fundamental symmetry breaking in the cavity field, it should generalise to a wide range of chiral systems. Our findings thus establish a viable paradigm for ultrafast enantiosensitive spectroscopy based on cavity quantum electrodynamics. More broadly, they illustrate how tailoring the photonic environment can amplify and control chiral molecular responses , which opens new avenues for detecting molecular handedness and for the light-matter control of chiral dynamics.

%%%%%%%%%%%%%%%%%%%%%%%%%%%%%%%%%%%%%%%%%%%%%%%%%%%%%%%%%%
{\em Acknowledgements:}  
%%%%%%%%%%%%%%%%%%%%%%%%%%%%%%%%%%%%%%%%%%%%%%%%%%%%%%%%%%
%\begin{acknowledgments} 
This work was supported by the National Key Research and Development Program of China (Grant No.\ 2024YFA1409800), NSFC Grant with No.\ 12274247, No.\ 12404310 and 12504279, Yongjiang talents program with No.\ 2022A-094-G and 2023A-158-G, Ningbo International Science and Technology Cooperation with No.\ 2023H009, `Lixue+' Innovation Leading Project and the foundation of national excellent young scientist. The Next Generation Chemistry theme at the Rosalind Franklin Institute is supported by the EPSRC (V011359/1 (P)) (A.J.). 

Author contributions 
H.G.D. convinced the research project. Y.-C.Y.\ performed the derivation and calculations. F.Z. supervised the {\em ab-initio} calculations. A.J. and H.G.D. supervised the project. A.J. and H.G.D. wrote the initial draft and all authors refined the final version of manuscript. 

%\end{acknowledgments}

%%%%%%%%%%%%%%%%%%%%%%%%%%%%%%%%%%%%%%%%%%%%%%%%%%%%%%%%%% Bibliography
%%%%%%%%%%%%%%%%%%%%%%%%%%%%%%%%%%%%%%%%%%%%%%%%%%%%%%%%%%
%\bibliographystyle{model1a-num-names}
%\bibliography{RefsFMO}

%%%%%%%%%%%%%%%%%%%%%%%%%%%%%%%%%%%%%%%%%%%%%%%%%%%%%%%%%%

\end{document}